\documentstyle[12pt,epsf]{article}

\newcommand{\beq}{\begin{equation}}
\newcommand{\eeq}{\end{equation}}
\newcommand{\bea}{\begin{eqnarray}}
\newcommand{\eea}{\end{eqnarray}}
\newcommand{\rar}{\rightarrow}
\newcommand{\lra}{\longrightarrow}
\newcommand{\lan}{\langle}
\newcommand{\ran}{\rangle}

\newcommand{\bk}{{\bf k}}
\newcommand{\bN}{{\bf N}}
\newcommand{\bY}{{\bf Y}}
\newcommand{\bone}{{\bf 1}}
\newcommand{\balpha}{\mbox{\boldmath$\alpha$}}
\newcommand{\bsigma}{\mbox{\boldmath$\sigma$}}
\newcommand{\cL}{{\cal L}}

\newcommand{\cH}{{\cal H}}
\newcommand{\bcN}{\mbox{\boldmath$\cal N$}}
\newcommand{\stheta}{S_{\theta}}
\newcommand{\ctheta}{C_{\theta}}
\newcommand{\sphi}{S_{\phi}}
\newcommand{\cphi}{C_{\phi}}

\begin{document}

\font\fortssbx=cmssbx10 scaled \magstep2
\hbox to \hsize{
\includegraphics{uwlogo.ps}
\hskip.5in \raise.1in\hbox{\fortssbx University of Wisconsin - Madison}
\hfill$\vcenter{\hbox{\bf MADPH-95-883}
            \hbox{April 1995}}$ }
\vskip 2cm
\begin{center}
\Large
{\bf On the validity
of  the reduced Salpeter equation} \\
\vskip 0.5cm
\large
 M. G. Olsson and  Sini\v{s}a Veseli \\
\vskip 0.1cm
{\small \em Department of Physics, University of Wisconsin, Madison,
	\rm WI 53706} \\
\vspace*{+0.2cm}
 Ken Williams \\
{\small \em Continuous Electron Beam Accelerator Facility \\
	Newport News, VA 29606, USA \\
	and \\
\vspace*{-0.2cm}
Physics Department, Hampton University, Hampton, VA 29668}
\end{center}
\thispagestyle{empty}
\vskip 0.7cm

\begin{abstract}
We adapt a general method to solve both the full and reduced Salpeter
equations and systematically explore the conditions under which
these two equations give equivalent
results in meson dynamics. The effects of constituent mass,
angular momentum state, type of interaction, and the
nature of confinement are all considered in an effort
to clearly delineate the range of validity of the
reduced Salpeter approximations.
We find that for $J\not{\hspace*{-1.0mm}=}0$ the solutions are
strikingly similar for all constituent masses.
For zero angular momentum states the full and
reduced Salpeter equations give different results
for small quark mass especially with a large additive constant
coordinate space potential. We also show that
$\frac{1}{m}$ corrections
to heavy-light energy levels can be
accurately computed with the reduced equation.
\end{abstract}

\newpage

\section{Introduction}
The instantaneous Bethe-Salpeter equation, or Salpeter equation
\cite{bib:salpeter}, is by far the most commonly employed
relativistic wave equation  in meson models with fer\-mio\-nic
constituents. Until
recently, almost all explicit calculations have used
a simplified version known as the reduced Salpeter equation. The later
becomes identical to the full Salpeter equation if at least
one of the constituent masses is infinite.

The reduced Salpeter equation is of the standard eigenvalue
(hermitian) type
whereas the full equation is not. Its solutions are thus
algebraically and numerically simpler than that of the full equation.
For example, the reduced equation doesn't have negative energy solutions,
nor does it have solutions with zero norm, both of which exist
for the full Salpeter equation \cite{bib:lagae1,bib:munz3}.
More importantly, the reduced
equation has variationally stable solutions for a
wider range of kernel types than does the full equation
\cite{bib:parramore,bib:spaper}.
For example,
there are no variationally
stable solutions to the full Salpeter equation corresponding
to pure scalar confinement. The reduced Salpeter equation, on the
other hand, has well defined variationally stable solutions with
scalar confinement.
Also, the reduced equation is equivalent to the ``no-pair'' equation
\cite{bib:sucher} proposed to cure the ``continuum
dissociation'' problem in relativistic atomic physics.
There are therefore historical, practical,
and physical reasons for using the reduced equation.
We outline here the
conditions under which this can be done without sacrificing
accuracy.

In the real world  the constituent mass is never infinite, so one faces
a quantitative question as to the practical region of
validity of the
reduced Salpeter equation. Our results here establish that for
many purposes the reduced Salpeter equation is quite adequate
and one can take advantage. An analysis
involving heavy-light mesons with
$c$ or $b$ quarks, or $b\bar{b}$, $c\bar{c}$, or $s\bar{s}$ onia states,
does not incur serious error by using the reduced Salpeter equation.
It is only
for $J=0$ states and with small quark masses where there
can be significant
differences between the full and reduced Salpeter solutions.
Dynamical models involving light pseudo-scalar
states such as the $\pi$, $\eta$, or $K$ mesons can lead
to serious errors if the full Salpeter equation is not used.

Our analysis draws heavily upon previous work \cite{bib:spaper}
in which we have adopted Laga$\ddot{\rm e}$'s method
 \cite{bib:lagae1} to
investigate the nature of  full Salpeter solutions. Our
principal conclusion was that the only linearly confining potential
which  yields linear Regge trajectories and
has  variationally stable solutions
is a time component Lorentz vector. This confirms previous
work done for the equal mass case \cite{bib:parramore,bib:lagae2}.

In the present work we use the fact that stable solutions
exist for the time component vector confinement in order
to estimate  the range of applicability
of the reduced Salpeter equation.
The desirable properties of the time
component vector potential in the Salpeter equation
does not mean
that it should be used as a confinement potential, since
it yields wrong sign of the spin-orbit interaction, disagreeing
both with QCD and experiment.
We also compare
solutions to the full equation and its reduced version
for an  equal mixture of scalar
and time component vector confinement. This type of mixed
confining kernel has been recently used in \cite{bib:munz3}
for the investigation of weak decays of heavy mesons.
The vector confinement stabilizes the scalar confining part
up to the case of equal mixtures. Phenomenologically, the scalar
confining part is necessary to reduce
the $P$-wave spin-orbit splitting. For this
mixed  confinement case we also explicitly demonstrate
that the reduced Salpeter equation is adequate for the investigation
of the heavy-light systems, such as $D$ and $B$ mesons, as
well as for heavy onia. We also examine
the extent
to which $\frac{1}{m}$ corrections to heavy-light systems depend on
which wave equation is used. We find that the difference is negligible
even for D mesons.

In Section \ref{sec:rev} we adapt Laga$\ddot{\rm e}$'s
 formalism \cite{bib:lagae1} to the reduced Salpeter equation. Our
numerical results are contained in Section \ref{sec:res}
where we compare the full and reduced Salpeter solutions for
both onia and heavy-light mesons. Our conclusions
are summarized in Section  \ref{sec:con}.
In the Appendix \ref{app:eqs} we provide the complete
reduced Salpeter radial equations for the three Lorentz
type kernels, $\gamma^{0}\otimes \gamma^{0}$ [time component
vector],
$\bone \otimes \bone$ [scalar], and $\gamma^{\mu}\otimes \gamma_{\mu}$
[full vector].

\section{Reduced Salpeter equation}
\label{sec:rev}

Recently
 Laga$\ddot{\rm e}$ has proposed an elegant formalism
\cite{bib:lagae1}
for the reduction
of the full Salpeter equation to a system of equations involving only
radial wave functions. One of the
nice things about his method is that the transition
from  full to  reduced Salpeter equations can be
accomplished easily. In this section
we briefly sketch the main points of this formalism as adapted to the
reduced Salpeter equation.

We start from the Salpeter equation for a fermion-antifermion system
in the CM frame of the bound state,
\bea
\Phi(\bk)& =& \int \frac{d^{3}\bk'}{(2\pi)^{3}}
\left[ \frac{\Lambda_{+}^{1}(\bk)\gamma^{0}[V(\bk,\bk')\Phi(\bk')]
\gamma^{0}\Lambda_{-}^{2}(-\bk)}{M-E_{1}-E_{2}} \right.\nonumber \\
&-&
\left. \frac{\Lambda_{-}^{1}(\bk)\gamma^{0}[V(\bk,\bk')\Phi(\bk')]
\gamma^{0}\Lambda_{+}^{2}(-\bk)}{M+E_{1}+E_{2}}
\right]\ .
\label{eq:seq}
\eea
Here, $\Lambda_{\pm}^{i}$'s are the usual energy
projection operators, given by
\beq
\Lambda_{\pm}^{i} =\frac{E_{i}(\bk)\pm H_{i}(\bk)}{2E_{i}(\bk)}\ ,
\eeq
with $H_{i}$ being the generalized Dirac Hamiltonians,
\beq
H_{i}(\bk) = A_{i}(\bk)\balpha\cdot \hat{\bk} + B_{i}(\bk)\beta\ ,
\eeq
and $E_{i}(\bk) = \sqrt{A_{i}(\bk)^{2}+B_{i}(\bk)^{2}}$.
Again, we'll  consider  constituent quarks of masses $m_{i}$,
so that
\bea
A_{i}(\bk) &=&k\ ,\label{eq:defa}\\
B_{i}(\bk) &=&m_{i}\ ,\\
E_{i}(\bk)&=&\sqrt{m_{i}^{2}+\bk^{2}}\label{eq:defe}\ .
\eea
The formal product of $V\Phi$ in equation (\ref{eq:seq}) represents
the sum of scalar potentials $V_{i}$ and bilinear covariants,
\beq
V(\bk,\bk')\Phi(\bk') \lra \sum_{i} V_{i}(\bk,\bk')
G_{i} \Phi(\bk') G_{i}\ ,
\eeq
where the $G_{i}$'s are Dirac matrices.

The reduced Salpeter equation is obtained by dropping the second term
from (\ref{eq:seq}), and this is usually justified
for heavy-quark systems on the grounds that
\beq
\frac{M-E_{1}-E_{2}}{M + E_{1}+E_{2}}\ll 1\ .
\eeq
The resulting equation,
\beq
M\Phi(\bk) = (E_{1}+E_{2})\Phi(\bk)+
\int \frac{d^{3}\bk'}{(2\pi)^{3}}
 \Lambda_{+}^{1}(\bk)\gamma^{0}[V(\bk,\bk')\Phi(\bk')]
\gamma^{0}\Lambda_{-}^{2}(-\bk)\ ,
\label{eq:rseq}
\eeq
is a standard eigenvalue equation, and it has been used in
a number of studies of relativistic bound states
\cite{bib:cung,bib:jacobs,bib:gara}.

In order to apply Laga$\ddot{\rm e}$'s formalism \cite{bib:lagae1} to
the reduced Salpeter equation,  we multiply (\ref{eq:rseq}) by
$\gamma^{0}$, and define
\bea
\chi(\bk) &=& \Phi(\bk)\gamma^{0}\ ,\\
\Gamma_{i} &=& \gamma^{0}G_{i}\ ,
\eea
so that (\ref{eq:rseq}) becomes
\beq
M\chi = (E_{1}+E_{2})\chi+
\sum_{i} \int \frac{d^{3}\bk'}{(2\pi)^{3}}V_{i}(\bk-\bk')
 \Lambda_{+}^{1}\Gamma_{i}\chi'
\Gamma_{i}\Lambda_{-}^{2}\ ,
\label{eq:rseq2}
\eeq
where notation $f=f(\bk),\ f'=f(\bk')$ is employed.  $V_{i}(\bk-\bk')$
has the Fourier transform $V(r)$ in the case
of Lorentz vector kernel, and $-V(r)$ in the case of Lorentz
scalar kernel.

Using properties of projection operators, it can be easily shown
that the full Salpeter amplitude satisfies
\beq
\frac{H_{1}}{E_{1}}\chi + \chi \frac{H_{2}}{E_{2}}=0\ .\label{eq:scon}
\eeq
For the reduced equation, this constraint breaks into two
parts,
\bea
H_{1}\chi &=& E_{1}\chi\ ,\label{eq:rscon1}\\
\chi H_{2} &=& - E_{2}\chi\ .\label{eq:rscon2}
\eea
Taking these constraints  into account,
the norm of the reduced Salpeter amplitude
\cite{bib:smith,bib:murota,bib:yao}
can be written as
\beq
||\chi ||^{2}=\int \frac{d^{3}\bk '}{(2\pi)^{3}}
{\rm Tr}\left[\chi^{\dagger}\chi\right]\ ,\label{eq:norm}
\eeq
and is related to the normalization of bound states as
\beq
||\chi||^{2}=\frac{1}{(2\pi)^{3}} \lan B| B\ran\ .
\eeq
Using (\ref{eq:rseq2}) inside of (\ref{eq:norm}) one obtains
\beq
M||\chi||^{2} = \int \frac{d^{3}\bk}{(2\pi)^{3}}[E_{1}+E_{2}]\
{\rm Tr}[\chi^{\dagger}\chi] +
\sum_{i} \int \frac{d^{3}\bk}{(2\pi)^{3}} \int \frac{d^{3}\bk'}{(2\pi)^{3}}
V_{i}(\bk -\bk')\ {\rm Tr}[\chi^{\dagger}\Gamma_{i}\chi'\Gamma_{i}]\ .
\label{eq:var}
\eeq
This equation will be used for obtaining radial equations from the
variational principle as outlined in \cite{bib:lagae1}. It is interesting
to note that it has the same form for both full and reduced
Salpeter equations.

Now, in the case of the full Salpeter equation, one
 expands the  amplitude as
\beq
\chi=\cL_{0} + \cL_{i}\rho_{i} + \bcN_{0}\cdot \bsigma +
\bcN_{i}\cdot \rho_{i}\bsigma\ ,\label{eq:ampl}
\eeq
with 16 Hermitian matrices whose squares are unity ($1,\rho_{i},
\bsigma,\rho_{i}\bsigma$) defined in
\cite{bib:lagae1}. Using this decomposition, it is
 easily seen that constraint (\ref{eq:scon})
can be satisfied by expressing the 16 components of $\chi$ ($\cL$'s and
$\bcN$'s) in terms of eight functions ($L_{1},L_{2},\bN_{1},\bN_{2}$).
The correct form for $\cL$'s and
$\bcN$'s is given in \cite{bib:spaper}.
For the reduced Salpeter equation, both constraints
(\ref{eq:rscon1}) and (\ref{eq:rscon2})
can be simultaneously satisfied if $L_{1}=L_{2}\equiv L$
and $\bN_{1}=\bN_{2}\equiv \bN$.

Following
\cite{bib:yao}, we
obtain the radial equations by  expressing $L$
and $\bN$ in terms of spherical harmonics and vector
spherical harmonics \cite{bib:lagae3}, so that
\bea
L(\bk) &=& L(k)Y_{JM}(\hat{\bk})\ ,\\
\bN(\bk) &=& N_{-}(k)\bY_{-}(\hat{\bk}) +
N_{0}(k)\bY_{0}(\hat{\bk}) +
N_{+}(k)\bY_{+}(\hat{\bk})\ ,
\eea
where $\bY_{-}$, $\bY_{0}$, and $\bY_{+}$, stand for
$\bY_{JJ-1M}$, $\bY_{JJM}$, and $\bY_{JJ+1M}$, respectively.
We also introduce functions $n_{+}$ and $n_{-}$, defined as
\beq
\left[ \begin{array}{c}
       n_{+} \\
      n_{-}
	\end{array} \right]
= \left[\begin{array}{cc}
\mu & \nu \\
    -\nu & \mu  \end{array}\right]
\left[\begin{array}{c}
	N_{+} \\
      N_{-} \end{array}\right] ,
\label{revers}
\eeq
with
\beq
\mu = \sqrt{\frac{J}{2J+1}}\ ,\ \nu = \sqrt{\frac{J+1}{2J+1}}\ .
\eeq
Using these definitions inside  expressions for the $\cL$'s and
$\bcN$'s as given in \cite{bib:spaper},
together with  properties of spherical
and vector spherical harmonics, (\ref{eq:var}) can
be expressed in terms of radial wave functions only. Then
by taking variations with respect to
$L^{*}(k),N_{0}^{*}(k),n_{+}^{*}(k),$ and $n_{-}^{*}(k)$,
as explained in \cite{bib:lagae1}, one obtains the
set of coupled equations for the radial wave functions
of the reduced Salpeter amplitude. We summarize these
equations in  Appendix \ref{app:eqs} for the
kernels $\gamma^{0}\otimes\gamma^{0}$, $\bone \otimes \bone$
and $\gamma^{\mu}\otimes \gamma_{\mu}$.

\section{Numerical results}
\label{sec:res}

As outlined in Appendix B of \cite{bib:spaper}, one can solve
the system of radial equations
by expanding the
wave functions in terms of a complete set of
basis states, which depend on a variational parameter $\beta$,
and then truncating the expansion to a finite
number of basis states. In this way, a set of coupled radial
equations can be transformed into a matrix equation,
$\cH \psi = M\psi$. The eigenvalues $M$ of the matrix $\cH$ will depend
on $\beta$, and by looking for the extrema of $M(\beta)$, one
can find the bound state energies. If the calculation is stable,
increasing the number of basis states used will
decrease the dependence of the eigenvalues
on $\beta$. Regions of $\beta$ with
the same eigenvalues should emerge and  enlarge. For each
of the results discussed in the remainder of this paper we have verified
that this indeed occurs.

\subsection{Equal mass case with $\gamma^{0}\otimes \gamma^{0}$ kernel}

In Figure \ref{fig:rfmm} we compare  solutions of reduced
and full Salpeter
equations for equal mass systems with a pure
time component vector confinement
($V(r)=ar$, $a = 0.2\ GeV^{2}$). We have varied
the quark masses
($m_{1}=m_{2}\equiv m$)
from 0 to $1\ GeV$, solved both equations for all
$J=0,1,\ {\rm and\ } 2$  states  (which involves
all $S$, $P$, and most $D$ waves), and plotted
the difference between state mass and
rest mass of the two quarks. As one can see, the difference between
the two solutions is noticeable only for $J^{PC}=0^{-+}$
and $0^{++}$ states, and then only for very small quark masses.
For example, for zero quark masses the difference for the
$0^{-+}$ state is about $25\ MeV$, while already for
quark masses of $0.3\ GeV$ it is only $6\ MeV$. On the other hand,
for the $1^{--}$
state the difference between the two solutions is about $1\ MeV$
even for zero quark masses.
Another interesting thing to observe in Figure \ref{fig:rfmm}
is that for both equations and for zero quark mass
 we have degeneracy of
$0^{-+}$ and $0^{++}$,  $1^{--}$ and $1^{++}$, and also
$2^{++}$ and $2^{--}$
states. This parity degeneracy
can be easily explained by referring to the
radial equations for the full Salpeter equation given in
Appendix A of \cite{bib:spaper}. In the limit where both masses
go to zero, it can be easily seen that $0^{-+}$ and $0^{++}$
equations are the same. Similarly, for $J>0$ states
the four radial equations
for $P=(-1)^{J}$ and $C=(-1)^{J}$ (involving
$n_{1+},n_{2+},n_{1-}$ and $n_{2-}$) decouple into two systems
of two equations. The first one
(involving $n_{1+}$ and $n_{2+}$)
is the same as
the system describing $P=(-1)^{J+1}$ and $C=(-1)^{J+1}$ states, while
 the second one (involving $n_{1-}$ and $n_{2-}$)
is equivalent to  the system describing $P=(-1)^{J+1}$ and $C=(-1)^{J}$
states (and higher in energy, as can be seen in
Figure \ref{fig:rfmm}). The $m=0$ degeneracy is an example of the chiral
symmetry of the vector potential and
its Wigner-Weyl realization through parity doublets.

In order to see the effects of the short range Coulomb potential,
we have performed a similar analysis with $V(r)=ar-\frac{\kappa}{r}$,
using $a=0.2\ GeV^{2}$ and $\kappa=0.5$.
The results are shown in Figure \ref{fig:rfmm2}. Again,  the
difference between full and reduced Salpeter solutions is
noticeable only for the $J^{PC}=0^{-+}$
and $0^{++}$ states. For the $0^{-+}$ state,
the difference is now about $35\ MeV$ for $m_{1}=m_{2}=m=0$, and
about $10\ MeV$ for $m_{1}=m_{2}=m=0.3\ GeV$. For the $1^{--}$ state,
the difference is only about $3\ MeV$ for zero quark masses.

Finally, in Figure \ref{fig:rfmm3}  we show the results of  the
same analysis as above, but this time
 including an additive constant,
$V(r)=ar+C-\frac{\kappa}{r}$
($a=0.2\ GeV^{2}$, $C=-1.0\ GeV$, and $\kappa=0.5$).
One might expect that adding a constant to
the potential would not change the difference
between the two equations. However, as one can
see from the Figure \ref{fig:rfmm3},  it is not quite like that.
Now the solutions to the full Salpeter
equation for the $0^{-+}$ and $0^{++}$ states
are considerably lower in energy than the solutions to the
reduced Salpeter equation. For the $0^{-+}$ ($1^{--}$) state
the difference is about $106\ MeV$ ($7\ MeV$)
 with zero quark masses and about $3\ MeV$
($1\ MeV$) with $m_{1}=m_{2}=1.0\ GeV$.

 The reason for this somewhat unexpected behavior is that a
negative constant $C$ added to the
kernel of the full Salpeter equation lowers the eigenvalues by an
amount larger than $|C|$, while  for the
reduced Salpeter equation it is exactly $|C|$.
For example, adding $C=-1.0\ GeV$
to the potential $V(r)=ar-\frac{\kappa}{r}$ with $a=0.2\ GeV^{2}$
and $\kappa=0.5$ the lowest eigenvalue for the $0^{-+}$
state (with zero quark masses)
 is lowered by about $1.072\ GeV$ for the full Salpeter equation
with a time component vector kernel. This effect is much
less noticeable with larger quark masses, and
higher $J$ states, e.g. for the $1^{--}$
state with zero quark masses and
same $a$ and $\kappa$ as before, the lowest eigenvalue
was lowered by $1.004\ GeV$ after adding $C=-1.0\ GeV$.
We also note that these numerical results were obtained
with 25 basis states, so that dependence of the
results on the variational parameter characterizing
the basis states
was negligible.

In order to further explore the relationship
between the
 full Salpeter equation and the reduced one, we
have plotted the radial wave functions for the
$0^{-+}$ case and for $V(r)=ar+C-\frac{\kappa}{r}$
($a=0.2\ GeV^{2}$, $C=-1.0\ GeV$, and $\kappa=0.5$).
 Just as a reminder, the
reduced Salpeter equation for the pseudoscalar case
has only one wave function ($L$), as opposed to
two ($L_{1}$ and $L_{2}$) in the full equation. Also, when
the  reduced Salpeter equation is valid, then
$L_{1}$ and $L_{2}$ are equal. As we can see from Figure
\ref{fig:ff1}, for very small quark masses ($m_{1}=m_{2}=0$),
the difference between $L_{1}$ and $L_{2}$ is large, and the
reduced equation cannot  replace  the full one. However,
with $m_{1}=m_{2}=1.0\ GeV$ (Figure \ref{fig:ff2}), the reduced
Salpeter result is much more closer to the full one.
In these two figures we use a Cornell potential
with an additive constant ($a=0.2\ GeV^{2},\ C=-1.0\ GeV$,
and $\kappa = 0.5$).

{}From this analysis, it is  clear that the solutions of the
reduced Salpeter equation
are nearly the same as those of the full one
 for the description of the heavy-heavy
($c\bar{c}$ and $b\bar{b}$) mesons, and a very good
first approximation even for the $s\bar{s}$ mesons
(with $s$ quark mass of about $500\ MeV$). This justifies the
assumption of Gara et. al. \cite{bib:gara}  that the reduced
Salpeter equation could be used for the description of $s\bar{s}$ mesons.

\subsection{Heavy-light case with $\gamma^{0}\otimes \gamma^{0}$ kernel}

A similar analysis can be performed for the  ``heavy-light''
systems. For $V(r)=ar+C-\frac{\kappa}{r}$, with
$a=0.2\ GeV^{2}$, $C=-1.0\ GeV$, and $\kappa=0.5$,
we fixed the ``light'' quark mass at $m_{1}=0$,
varied the ``heavy'' antiquark mass $m_{2}$ from 0 to $1\ GeV$, and
solved both equations for all $J=0,1$ and $2$ states. The results
are shown in Figure \ref{fig:rfm}. The degeneracy
of states with the same $J$ and different parity
can be again explained easily by looking
into the radial equations
for the full Salpeter equation given in
Appendix A of \cite{bib:spaper}. In the limit where
$m_{1}\rar 0$, $\phi_{1}\rar 0$ and $\phi\rar \theta$,
which makes equivalent the two sets of equations for
different parities. As far as the difference
between the full and reduced Salpeter equations are concerned, it
is again important
only for $J=0$ states. For example, for $0^{-}$ and $0^{+}$
($1^{-}$ and $1^{+}$) it is only about $7\ MeV$ ($1\ MeV$) at
$m_{2}=1.0\ GeV$.
Figure \ref{fig:ff3} shows that for
$m_{1}=0,\ m_{2}=1\ GeV,$  $L$ is already a very good approximation
to $L_{1}$ and $L_{2}$.
One also has to remember that with such a large negative constant
the $c$ quark mass must be considerably
larger than $1.0\ GeV$
in order to describe $D$  mesons.
Given all this, we conclude that the reduced Salpeter equation is an
excellent approximation to the full one for the description
of $D$ and $B$ mesons.

Although the time component vector interaction has many nice properties,
it is flawed as a realistic quark confinement interaction.
As pointed out earlier, it predicts
``parity doubling'' of meson states in the limit of zero quark mass.
For large quark masses this difficulty appears as the ``wrong sign''
spin-orbit interaction which conflicts both with
experiment and QCD.

\subsection{Mixed confinement potentials}

As already mentioned, recently a half-half mixture of the time
component vector and scalar confinement has been
proposed in \cite{bib:munz3},
together with a one gluon exchange kernel, for the investigation
of  weak decays of $B$ and $D$ mesons. In order to compare the
full Salpeter equation with its reduced version in this type
of model, we adopt the mixed confining kernel,
\beq
\frac{1}{2}[ \gamma^{0}\otimes \gamma^{0} + \bone\otimes\bone]V_{c}(r)\ ,
\label{eq:mix1}
\eeq
with
\beq
V_{c}(r)=ar + C\ ,
\eeq
and for the short range potential we simply take
\beq
[\gamma^{0}\otimes \gamma^{0}]V_{g}(r)\ ,
\eeq
where
\beq
V_{g}(r)=-\frac{\kappa}{r}\ .
\label{eq:mix2}
\eeq
A confinement mixture of this type has been
shown to have a stable variational solution \cite{bib:spaper}.
For the parameters of the potential we choose
$a=0.2\ GeV^{2}$, $C=-1.0\ GeV$, and $\kappa=0.5$. Computation
of the equal mass
case
 is shown in Figure \ref{fig:v0smm}
(for the $0^{-+}$ and $0^{++}$ states). As one can see,
the differences between  full and reduced Salpeter
solutions are only slightly different than in the case
with a pure time component vector kernel. For the $0^{-+}$
state and $m_{1}=m_{2}=m=1.0\ GeV$ the
difference between the two equations is about $7\ MeV$.
The heavy-light case calculation (for the same potential parameters) is
shown in Figure \ref{fig:v0sm}. The difference between
 the two solutions for the $0^{-}$ state, and for
$m_{1}=0$ and $m_{2}=1.0\ GeV$, is about $9\ MeV$.
Therefore, we again conclude that the reduced
Salpeter equation is  as good as the full
Salpeter equation for the description
of the $c\bar{c}$ and $b\bar{b}$ mesons,
 a very good first approximation even for the $s\bar{s}$ mesons,
and would serve  as well as the full
Salpeter equation for the description of the heavy-light
systems, such as $D$ and $B$ mesons.

Of course, these results are dependent on  parameters of the
particular model. However, in our analysis we have used
values for $a$ and $\kappa$ that are
typical in the hadron spectroscopy, and constant
$C$ that is perhaps  slightly larger than usual. We have
also  restricted ourselves to constituent masses that are
smaller than the usually assumed  $c$ quark mass.
Therefore, we feel that our main conclusions would not
be drastically altered if a different set of realistic parameters was used.

In order to illustrate this, we have chosen
 parameters of the potential to be as close as possible
 to the ones used
in \cite{bib:munz3} (as given in their Table 1), i.e.
$m_{1}=0.2\ GeV$, $m_{2}=1.738\ GeV$, $a=0.335\ GeV^{2}$, $C=-1.027\ GeV$,
and $\kappa =0.521$ (which corresponds to $\alpha_{sat}=0.391$
in \cite{bib:munz3}), and solved both equations for the $0^{-}$
and $0^{+}$ states, with the
kernel described by (\ref{eq:mix1}-\ref{eq:mix2}). The
differences between ground state energies were $5\ MeV$ and
$0\ MeV$, respectively, despite
the large value of $a$.
 For the $0^{-}$ state, where the
difference between the two solutions should be most obvious,
 we have plotted the radial wave
 functions in Figure \ref{fig:ff4}. As one can see, the
reduced wave function is a very good
approximation for the full wave functions.

For the sake of simplicity, in the previous
calculations we have used a short range
potential with a fixed coupling constant, for which Murota
\cite{bib:murota} has shown
 most of the Salpeter amplitudes are divergent as
$r\rar 0$. If one uses a running coupling constant, this
divergence is less pronounced, but still present.
That is
precisely the reason why
 the short range potential used in \cite{bib:munz3} was
regularized.
In order to show the effects of regularization, instead
of (\ref{eq:mix2}) we now take as  in \cite{bib:munz3}
\beq
V_{g}(r)=\left\{
\begin{array}{cc}
-\frac{4}{3}\frac{\alpha(r)}{r}\ , & r\geq r_{0} \\
a_{g}r^{2}+b_{g}\ , & r< r_{0}
\end{array}
\right. \ .
\label{eq:rcoul}
\eeq
The constants
$a_{g}$ and $b_{g}$ are determined
by the condition that $V_{g}(r)$ and its derivative are
continuous functions. The running coupling constant is
parametrized exactly as in
 \cite{bib:munz3}, with their value of
 $r_{0}=0.507\ GeV^{-1}$, and the saturation  value
for the coupling constant  $\alpha_{sat}=0.391$. The string
tension and constant were again $a=0.335\ GeV^{2}$ and
$C=-1.027\ GeV$, and
 quark masses
were $m_{1}=0.2\ GeV$ and $m_{2}=1.738\ GeV$, as for the previous
calculation.
Using these parameters, we have
again solved both equations for the $0^{-}$ and
$0^{+}$ states. The
differences between ground state energies were
$3\ MeV$ and
$0\ MeV$, respectively, showing that
a regularized short range potential reduces
the differences between the reduced and the full Salpeter
equation.
 For the $0^{-}$ state,
 we have again plotted the radial wave
 functions in Figure \ref{fig:ff5}. As one can see, all
wave functions are now finite at the origin, and the
reduced Salpeter wave function is  an even better
approximation to the full ones than it was before.

We can also use this model to estimate the accuracy of $\frac{1}{m}$
recoil
corrections to the heavy-light limit. In Figure \ref{fig:mcorr}
we show the difference between the $0^{-}$
ground states for a finite and an infinite heavy
mass ($m_{2}$) with a massless light quark ($m_{1}$) in both cases.
We see that these ``recoil'' corrections are quite important
even for the $b$ quark mesons where correction is nearly $40\ MeV$
(at $\frac{1}{m_{2}}\simeq 0.2 \ GeV^{-1}$). On the other hand,
the difference between full and reduced Salpeter solutions is small.
 For a charmed meson ($\frac{1}{m_{2}}\simeq 0.66 \ GeV^{-1}$) the
difference is about $3.5\ MeV$, while for a meson with a $b$ quark
it is about $0.2\ MeV$.

In \cite{bib:munz3} the mixed confinement (\ref{eq:mix1})
was used in part because the full Salpeter
equation does not have stable solutions unless the
scalar confinement part is equal or less  than the
time component vector part. We should note that the pure scalar confinement
could have been used with the reduced Salpeter equation.

\section{Conclusions}
\label{sec:con}

The reduced Salpeter equation, also known as the no-pair equation,
has long been used in dynamical
models of mesons. It has also
been long appreciated that it is an approximation
to the full Salpeter equation and that the discarded portion
 only vanishes if at least one of the constituent  masses is
infinite.
The reduced
equation has nevertheless been used because it has the standard
hermitian form.

In this paper we have examined the conditions under which
the reduced equation can be employed
without significant loss in accuracy. The critical factors
turn out to be constituent mass, $J^{P}$ state, and
the nature of the interaction.
If the total quark mass exceeds about $1.0\ GeV$ very little difference
is found between the full and reduced Salpeter solutions. Also,
with the exception of the $0^{-}$ and
$0^{+}$ states very little difference is found even at
 zero quark mass.
Finally, even
for $0^{\pm}$ states and vanishing quark mass the differences
between full and reduced Salpeter solutions are small if
there is no large constant in the coordinate space
confining potential.

There remain a number of hadronic states with light quark masses
 in which
the full Salpeter equation must be used. Differences
up to $100\ MeV$ were found between pseudoscalar
masses at zero quark mass for the two equations.

In our comparison between the full and reduced
Salpeter solutions we have considered both energies
and wave functions. As was the case with the energy eigenvalues,
we see large differences between the $0^{-}$ full and reduced
wave functions for zero quark mass (see Fig. \ref{fig:ff1}).
The differences are largest at the origin, $r=0$.
As observed in subsequent figures,
increasing the quark mass and considering higher states
causes the reduced Salpeter wave functions to become more
similar to the full ones. The difference between the two solutions
is always most noticeable at the origin.

We have primarily considered the time component vector
kernel, since its solutions with the full Salpeter
equation are variationally stable and yield normal
linear Regge trajectories in the case of linear confinement
\cite{bib:spaper}.
Although the solutions with a time component vector
potential have many desirable properties, a quark confinement
of this type has a spin-orbit interaction
of the wrong sign.
The addition of up to equal parts Lorentz
scalar confinement has been advocated recently \cite{bib:munz3}
for the study of weak decays of heavy-light mesons. The variational
stability is retained in this case and the reduced Salpeter equation
is shown to be accurate under similar conditions as in
the pure time component vector case. The reduced equation
has the additional advantage of variational stability with pure scalar
confinement.

\newpage

\appendix

\begin{center}
\vskip 0.5cm
APPENDIX
\end{center}

\section{Radial equations}
\label{app:eqs}

In this appendix we give the final form of the radial equations
for the reduced Salpeter equation
for the
kernels $\gamma^{0}\otimes\gamma^{0}$, $\bone \otimes \bone$
and $\gamma^{\mu}\otimes \gamma_{\mu}$.
These equations represent
a general case with a quark of mass $m_{1}$ and an
anti-quark of mass $m_{2}$.
However, one has to keep in mind
that for $J=0$ two wave functions vanish, i.e.
 we have $N_{0}=0$,
and $n_{+}=0$.

As in \cite{bib:spaper,bib:lagae1} we have used notation
\bea
\sphi = \sin{\phi}\ ,\ \cphi=\cos{\phi}\ ,\\
\stheta = \sin{\theta}\ ,\ \ctheta=\cos{\theta}\ ,
\eea
with angles $\phi$ and $\theta$ defined as
\beq
\phi=\frac{\phi_{1}+\phi_{2}}{2}\ ,\
\theta=\frac{\phi_{2}-\phi_{1}}{2}\ ,
\eeq
while $\phi_{i}$'s are defined through
\beq
\cos{\phi_{i}}=\frac{A_{i}}{E_{i}}\ ,\
\sin{\phi_{i}}=\frac{B_{i}}{E_{i}}\ .
\eeq
$A_{i}$, $B_{i}$ and $E_{i}$ are defined in (\ref{eq:defa}-\ref{eq:defe}).

In the equal mass case the equations given
below somewhat simplify, since
one has $E_{1}=E_{2}$, $\phi=\phi_{1}=\phi_{2}$,
and $\theta = 0$, so that
$\stheta=0$ and $\ctheta=1$. Also, since charge conjugation
 is a good quantum number
in the equal mass case, the two $P=(-1)^{J+1}$ state
equations decouple into two separate equations, one corresponding
to $C=(-1)^{J}$ (involving $L$), and the other
corresponding to $C=(-1)^{J+1}$ (involving $N_{0}$).

The heavy-light limit
($m_{2}\rar \infty$) is obtained by setting $E_{2}\rar m_{2}$,
$\phi_{2}\rar \frac{\pi}{2}$, so that $\stheta\rar \cphi$ and
$\ctheta\rar \sphi$. As expected, in the heavy-light
limit equations for the $\gamma^{0}\otimes\gamma^{0}$ and
 $\gamma^{\mu}\otimes\gamma_{\mu}$ kernels are the same.
Also, in this case spin of
the heavy  quark decouples from the spin of the light quark,
so that total angular momentum $j$ of the light quark becomes
a good quantum number. Inverting (\ref{revers}),
\bea
N_{+} &=& \mu n_{+} - \nu n_{-}\ ,\\
N_{-} &=& \nu n_{+} + \mu n_{-}\ ,
\eea
and also putting
\bea
L_{+} &=& \nu L - \mu N_{0}\ ,\\
L_{-} &=& \mu L + \nu N_{0}\ ,
\eea
from the heavy-light limit equations in terms of
$n_{+}$, $n_{-}$, $N_{0}$ and $L$, one can obtain
decoupled equations in terms of
$N_{+}$, $N_{-}$, $L_{+}$ and $L_{-}$, describing heavy-light
states with quantum number $j$. There will always be
a pair of degenerate states, described with
$N_{-}$ and $L_{+}$ ($J=L+1$ and $J=L$, for the
state with  $j=L+\frac{1}{2}$),
and
$N_{+}$ and $L_{-}$ ($J=L-1$ and $J=L$,
for the state with with $j=L-\frac{1}{2}$).

 For any mixture
of different kernels, only the
kernel parts of the radial equations should
be added. The kinetic energy terms are always the same. In the
$\bone\otimes \bone$ case, we have introduced an
additional minus sign
in the kernel, so that
$V(r)$ has the same form for all three cases considered, e.g. for the
Cornell potential $V(r) = a r -\frac{\kappa}{r}$.

\subsection{$\gamma^{0}\otimes\gamma^{0}$ kernel}

States with parity $P=(-1)^{J+1}$:
\bea
M L &=& [E_{1}+E_{2}] L
+ \frac{1}{2}\int_{0}^{\infty} \frac{k'^{2}dk'}{(2\pi)^{2}}
[\ctheta V_{J} \ctheta' L' +\sphi V_{J} \sphi' L' \nonumber \\
&+&\stheta (\mu^{2}V_{J-1}+\nu^{2}V_{J+1})\stheta'L'
+\cphi (\mu^{2}V_{J-1}+\nu^{2}V_{J+1})\cphi'L' \nonumber \\
&+&\mu\nu \stheta (V_{J-1}-V_{J+1})\cphi' N_{0}'
+\mu\nu \cphi (V_{J-1}-V_{J+1})\stheta' N_{0}'] \ ,\\
M N_{0} &=& [E_{1}+E_{2}] N_{0}
+ \frac{1}{2}\int_{0}^{\infty}
 \frac{k'^{2}dk'}{(2\pi)^{2}} [\ctheta V_{J} \ctheta'N_{0}'
+\sphi V_{J} \sphi' N_{0}' \nonumber \\
&+&\cphi (\nu^{2}V_{J-1}+\mu^{2}V_{J+1})\cphi'N_{0}'
+\stheta (\nu^{2}V_{J-1}+\mu^{2}V_{J+1})\stheta'N_{0}'\nonumber \\
&+&\mu\nu \cphi (V_{J-1}-V_{J+1})\stheta' L'
+\mu\nu \stheta (V_{J-1}-V_{J+1})\cphi' L'] \ .\nonumber
\eea

States with parity $P=(-1)^{J}$:
\bea
M n_{+} &=& [E_{1}+E_{2}] n_{+}
+ \frac{1}{2}\int_{0}^{\infty}
 \frac{k'^{2}dk'}{(2\pi)^{2}} [\cphi V_{J} \cphi' n_{+}'
+\stheta V_{J} \stheta' n_{+}'
\nonumber \\
&+&\sphi (\nu^{2}V_{J-1}+\mu^{2}V_{J+1})\sphi'n_{+}'
+\ctheta (\nu^{2}V_{J-1}+\mu^{2}V_{J+1})\ctheta'n_{+}'\nonumber \\
&+&\mu\nu \sphi (V_{J-1}-V_{J+1})\ctheta' n_{-}'
+\mu\nu \ctheta (V_{J-1}-V_{J+1})\sphi' n_{-}'] \ ,\\
M n_{-} &=& [E_{1}+E_{2}] n_{-}
+ \frac{1}{2}\int_{0}^{\infty}
 \frac{k'^{2}dk'}{(2\pi)^{2}} [
\cphi V_{J} \cphi' n_{-}'+\stheta V_{J} \stheta' n_{-}'
\nonumber \\
&+&\ctheta (\mu^{2}V_{J-1}+\nu^{2}V_{J+1})\ctheta'n_{-}'
+\sphi (\mu^{2}V_{J-1}+\nu^{2}V_{J+1})\sphi'n_{-}'\nonumber\\
&+&\mu\nu \ctheta (V_{J-1}-V_{J+1})\sphi' n_{+}'
+\mu\nu \sphi (V_{J-1}-V_{J+1})\ctheta' n_{+}'] \ .\nonumber
\eea

\subsection{$\bone\otimes \bone$ kernel}

States with parity $P=(-1)^{J+1}$:
\bea
M L &=& [E_{1}+E_{2}] L
+ \frac{1}{2}\int_{0}^{\infty}
 \frac{k'^{2}dk'}{(2\pi)^{2}}
[\ctheta V_{J} \ctheta' L' +\sphi V_{J} \sphi' L'
\nonumber \\
&-&\stheta (\mu^{2}V_{J-1}+\nu^{2}V_{J+1})\stheta'L'
-\cphi (\mu^{2}V_{J-1}+\nu^{2}V_{J+1})\cphi'L'
\nonumber \\
&-&\mu\nu \stheta (V_{J-1}-V_{J+1})\cphi' N_{0}'
-\mu\nu \cphi (V_{J-1}-V_{J+1})\stheta' N_{0}'] \ ,\\
M N_{0} &=& [E_{1}+E_{2}] N_{0}
+ \frac{1}{2}\int_{0}^{\infty}
 \frac{k'^{2}dk'}{(2\pi)^{2}} [\ctheta V_{J} \ctheta'N_{0}'
+\sphi V_{J} \sphi' N_{0}'
\nonumber \\
&-&\cphi (\nu^{2}V_{J-1}+\mu^{2}V_{J+1})\cphi'N_{0}'
-\stheta (\nu^{2}V_{J-1}+\mu^{2}V_{J+1})\stheta'N_{0}'\nonumber \\
&-&\mu\nu \cphi (V_{J-1}-V_{J+1})\stheta' L']
-\mu\nu \stheta (V_{J-1}-V_{J+1})\cphi' L'] \ .\nonumber
\eea

States with parity $P=(-1)^{J}$:
\bea
M n_{+} &=& [E_{1}+E_{2}] n_{+}
+ \frac{1}{2}\int_{0}^{\infty}
 \frac{k'^{2}dk'}{(2\pi)^{2}} [-\cphi V_{J} \cphi' n_{+}'
-\stheta V_{J} \stheta' n_{+}'
\nonumber \\
&+&\sphi (\nu^{2}V_{J-1}+\mu^{2}V_{J+1})\sphi'n_{+}'
+\ctheta (\nu^{2}V_{J-1}+\mu^{2}V_{J+1})\ctheta'n_{+}'\nonumber \\
&+&\mu\nu \sphi (V_{J-1}-V_{J+1})\ctheta' n_{-}'
+\mu\nu \ctheta (V_{J-1}-V_{J+1})\sphi' n_{-}'] \ ,\\
M n_{-} &=& [E_{1}+E_{2}] n_{-}
+ \frac{1}{2}\int_{0}^{\infty}
 \frac{k'^{2}dk'}{(2\pi)^{2}} [
-\cphi V_{J} \cphi' n_{-}' -\stheta V_{J} \stheta' n_{-}'
\nonumber \\
&+&\ctheta (\mu^{2}V_{J-1}+\nu^{2}V_{J+1})\ctheta'n_{-}'
+\sphi (\mu^{2}V_{J-1}+\nu^{2}V_{J+1})\sphi'n_{-}'\nonumber\\
&+&\mu\nu \ctheta (V_{J-1}-V_{J+1})\sphi' n_{+}'
+\mu\nu \sphi (V_{J-1}-V_{J+1})\ctheta' n_{+}'] \ .\nonumber
\eea

\subsection{$\gamma^{\mu}\otimes\gamma_{\mu}$ kernel}

States with parity $P=(-1)^{J+1}$:
\bea
M L &=& [E_{1}+E_{2}] L
+ \int_{0}^{\infty}
 \frac{k'^{2}dk'}{(2\pi)^{2}}
[2\ctheta V_{J} \ctheta' L' -\sphi V_{J} \sphi' L'
\nonumber \\
&+&\stheta (\mu^{2}V_{J-1}+\nu^{2}V_{J+1})\stheta'L'
+\mu\nu \stheta (V_{J-1}-V_{J+1})\cphi' N_{0}'
 \ ,\\
M N_{0} &=& [E_{1}+E_{2}] N_{0}
+ \int_{0}^{\infty}
 \frac{k'^{2}dk'}{(2\pi)^{2}} [\ctheta V_{J} \ctheta'N_{0}'
\nonumber \\
&+&\cphi (\nu^{2}V_{J-1}+\mu^{2}V_{J+1})\cphi'N_{0}'
+\mu\nu \cphi (V_{J-1}-V_{J+1})\stheta' L']
\ .\nonumber
\eea

States with parity $P=(-1)^{J}$:
\bea
M n_{+} &=& [E_{1}+E_{2}] n_{+}
+ \int_{0}^{\infty}
 \frac{k'^{2}dk'}{(2\pi)^{2}} [\cphi V_{J} \cphi' n_{+}'
\nonumber \\
&+&\ctheta (\nu^{2}V_{J-1}+\mu^{2}V_{J+1})\ctheta'n_{+}'
+\mu\nu \ctheta (V_{J-1}-V_{J+1})\sphi' n_{-}'] \ ,\\
M n_{-} &=& [E_{1}+E_{2}] n_{-}
+ \int_{0}^{\infty}
 \frac{k'^{2}dk'}{(2\pi)^{2}} [2\cphi V_{J} \cphi' n_{-}'
-\stheta V_{J} \stheta' n_{-}'
\nonumber \\
&+&
\sphi (\mu^{2}V_{J-1}+\nu^{2}V_{J+1})\sphi'n_{-}'
+\mu\nu \sphi (V_{J-1}-V_{J+1})\ctheta' n_{+}'] \ .\nonumber
\eea

\vskip 1cm
\begin{center}
ACKNOWLEDGMENTS
\end{center}
This work was supported in part by the U.S. Department of Energy
under Contract Nos.  DE-FG02-95ER40896 and DE-AC05-84ER40150,
the National Science
Foundation under Grant No. HRD9154080,
and in part by the University
of Wisconsin Research Committee with funds granted by the Wisconsin Alumni
Research Foundation.

\newpage

\begin{figure}
\begin{center}
FIGURES
\end{center}
\vskip 0.2cm
\end{figure}

\begin{figure}
\caption{Equal mass comparison of the reduced (dashed lines) and full
Salpeter (full lines) equations for the time component vector kernel with
$V(r)=ar$ ($a=0.2\ GeV^{2}$).
 The energies of all states with
$J$ equal to 0, 1, and 2 are shown as a function
of the quark mass. We have used 15 basis states.}
\label{fig:rfmm}
\end{figure}

\begin{figure}
\caption{Equal mass comparison of solutions
to the  reduced (dashed lines) and full
Salpeter (full lines) equations for the time component vector kernel with
$V(r)=ar-\frac{\kappa}{r}$ ($a=0.2\ GeV^{2},\kappa=0.5$).
The energies of all states with
$J$ equal to 0, 1, and 2 are shown as a function
of the quark mass. We have used 15 basis states.}
\label{fig:rfmm2}
\end{figure}

\begin{figure}
\caption{Equal mass comparison of the reduced (dashed lines) and full
Salpeter (full lines) equations for the time component vector kernel
with $V(r)=ar+C-\frac{\kappa}{r}$
($a=0.2\ GeV^{2}$, $C=-1.0\ GeV$, $\kappa=0.5$).
The energies of all states with
$J$ equal to 0, 1, and 2 are shown as a function
of the quark mass. We have  used 15 basis states.}
\label{fig:rfmm3}
\end{figure}

\begin{figure}
\caption{Pseudoscalar ($J^{PC}=0^{-+}$)
radial wave functions in  coordinate
space for the reduced ($L$, dashed line) and full Salpeter
equations ($L_{1}$,
lower full line, and $L_{2}$, upper full line), with
time component vector kernel and $V(r)=ar+C-\frac{\kappa}{r}$
($a=0.2\ GeV^{2}$, $C=-1.0\ GeV$, $\kappa=0.5$).
The quark masses were $m_{1}=m_{2}=0$, and the calculation was
done with 25 basis states.}
\label{fig:ff1}
\end{figure}

\begin{figure}
\caption{Pseudoscalar ($J^{PC}=0^{-+}$)
radial wave functions in  coordinate
space for the reduced ($L$, dashed line) and full Salpeter
equations ($L_{1}$,
lower full line, and $L_{2}$, upper full line), with
time component vector kernel and $V(r)=ar+C-\frac{\kappa}{r}$
($a=0.2\ GeV^{2}$, $C=-1.0\ GeV$, $\kappa=0.5$).
The quark masses $m_{1}=m_{2}=1\ GeV$, and the calculation was
done with 25 basis states.}
\label{fig:ff2}
\end{figure}

\begin{figure}
\caption{Comparison for heavy-light mesons
of the reduced (dashed lines) and full
Salpeter (full lines) solutions for the time component vector kernel
with $V(r)=ar+C-\frac{\kappa}{r}$ ($a=0.2\ GeV^{2}$, $C=-1.0\ GeV$,
$\kappa=0.5$). The lighter quark mass was fixed
at $m_{1}=0$, and we show the light degree
of freedom energy $M-m_{2}$ as a function
of $m_{2}$ for the lowest
angular momentum states $J^{P}$. We have used 15 basis states.}
\label{fig:rfm}
\end{figure}

\begin{figure}
\caption{Heavy-light pseudoscalar ($J^{P}=0^{-}$)
radial wave functions in  coordinate
space for the reduced ($L$, dashed line) and full Salpeter
equations ($L_{1}$,
lower full line, and $L_{2}$, upper full line), with
time component vector kernel and $V(r)=ar+C-\frac{\kappa}{r}$
($a=0.2\ GeV^{2}$, $C=-1.0\ GeV$, $\kappa=0.5$).
The quark masses were $m_{1}=0$ and $m_{2}=1\ GeV$. The calculation was
done with 25 basis states.}
\label{fig:ff3}
\end{figure}

\begin{figure}
\caption{Equal mass comparison of the reduced (dashed lines) and full
Salpeter (full lines) ground state
$0^{-+}$ and $0^{++}$ energies. An equal
mixture of the time component vector and scalar confinement
($V_{c}(r)=ar+C$), together with time component
vector short range potential ($V_{g}(r)=-\frac{\kappa}{r}$) was used.
The potential parameters were
$a=0.2\ GeV^{2}$, $C=-1.0\ GeV$, and $\kappa=0.5$.
Comparison with Figure \protect\ref{fig:rfmm3}
shows the breaking of the parity degeneracy at $m=0$.
15 basis states was used for calculation.}
\label{fig:v0smm}
\end{figure}

\begin{figure}
\caption{Heavy-light mixed confinement
comparison of the reduced (dashed lines) and full
Salpeter (full lines) ground state
$0^{-}$ and $0^{+}$ energies. An equal
mixture of the time component vector and scalar confinement
($V_{c}(r)=ar+C$), together with time component
vector short range potential ($V_{g}(r)=-\frac{\kappa}{r}$) was used.
The potential parameters were
$a=0.2\ GeV^{2}$, $C=-1.0\ GeV$, and $\kappa=0.5$, while
the lighter constituent mass was fixed at $m_{1}=0$.
By comparing with Figure \protect\ref{fig:rfm} we observe
the lifting of
the
parity degeneracy present in a pure time component vector
interaction at $m=0$.
15 basis states was used in the calculation.}
\label{fig:v0sm}
\end{figure}

\begin{figure}
\caption{Pseudoscalar ($J^{P}=0^{-}$)
radial wave functions in  coordinate
space for the reduced ($L$, dashed line) and full Salpeter
equations ($L_{1}$,
lower full line, and $L_{2}$, upper full line), with
a half-half
mixture of the time component vector and scalar confinement
($V_{c}(r)=ar+C$), together with time component
vector short range potential ($V_{g}(r)=-\frac{\kappa}{r}$).
The potential parameters were
$a=0.335\ GeV^{2}$, $C=-1.027\ GeV$, and $\kappa=0.521$, while
the quark masses were $m_{1}=0.2$ and $m_{2}=1.738\ GeV$.
The calculation was
done with 25 basis states, and represents
a model of \protect\cite{bib:munz3}, but with
a singular short range
potential. }
\label{fig:ff4}
\end{figure}

\begin{figure}
\caption{Pseudoscalar ($J^{P}=0^{-}$)
radial wave functions in  coordinate
space for the reduced ($L$, dashed line) and full Salpeter
equations ($L_{1}$,
lower full line, and $L_{2}$, upper full line), with
a half-half
mixture of the time component vector and scalar confinement
($V_{c}(r)=ar+C$), together with the regularized time component
vector short range potential (as defined in
(\protect\ref{eq:rcoul}) in the text).
The potential parameters were
$a=0.335\ GeV^{2}$, $C=-1.027\ GeV$, $\alpha_{sat}=0.391$, and
$r_{0}=0.507\ GeV^{-1}$, while
quark masses were $m_{1}=0.2$ and $m_{2}=1.738\ GeV$.
The calculation was
done with 25 basis states, and represents
a model of \protect\cite{bib:munz3}, including
a regularized short range
potential.}
\label{fig:ff5}
\end{figure}

\begin{figure}
\caption{$\frac{1}{m}$ corrections to the heavy-light $0^{-}$
ground state energy as a function of $\frac{1}{m_{2}}$ using the model of
\protect\cite{bib:munz3}, with light
quark mass $m_{1}=0$, and potential parameters the same
as before. The correction
ranges from about $40\ MeV$ for the
$B$ meson to about $100\ MeV$ for the $D$ meson. The difference
between the full and reduced
solutions is $0.2\ MeV$ and $3.5\ MeV$ for the $B$ and $D$ mesons
respectively.}
\label{fig:mcorr}
\end{figure}

\begin{figure}
\vspace*{+0.1cm}
\end{figure}

\clearpage

\begin{figure}[p]
\epsfxsize = 5.4in \epsfbox{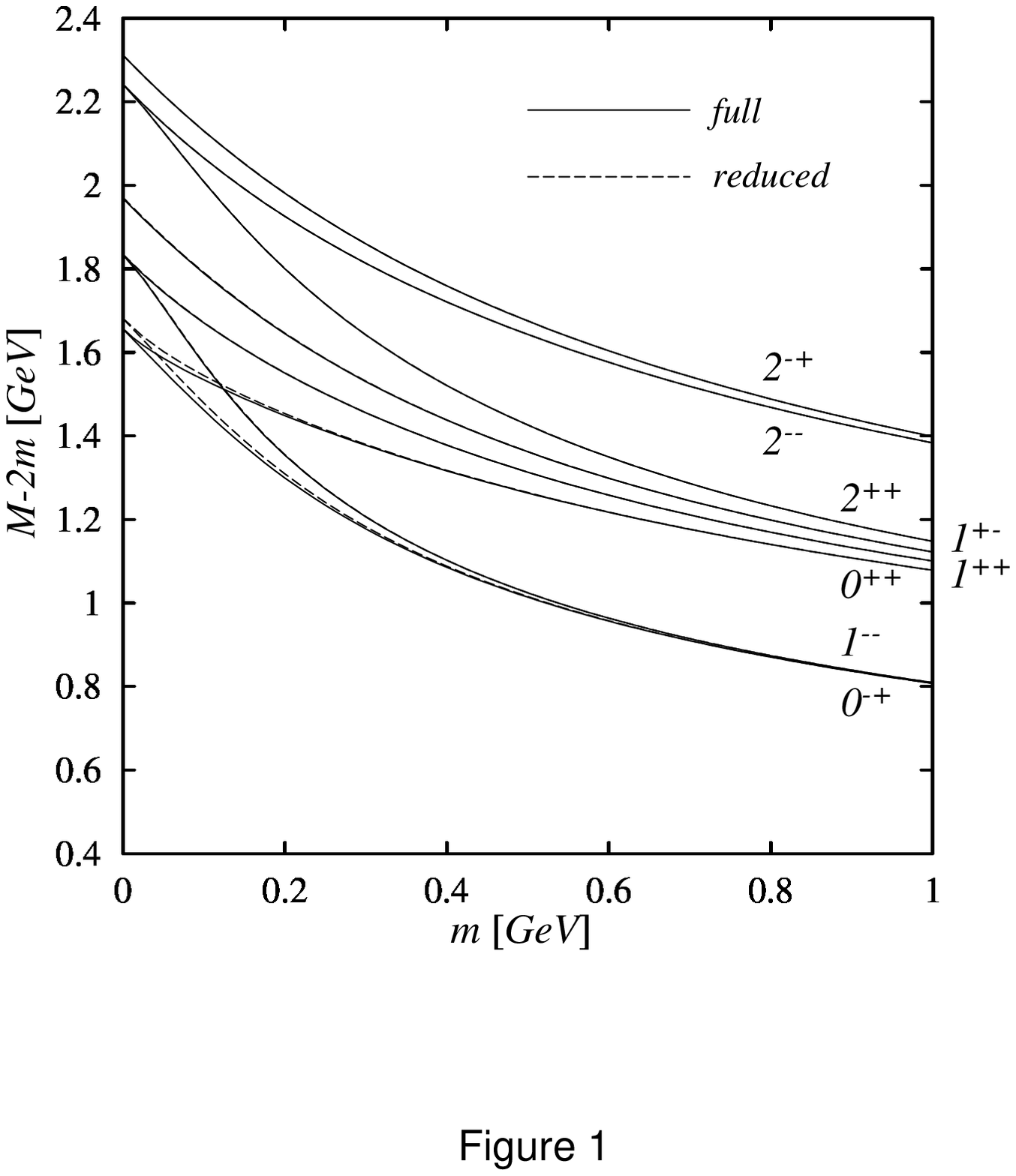}
\end{figure}

\begin{figure}[p]
\epsfxsize = 5.4in \epsfbox{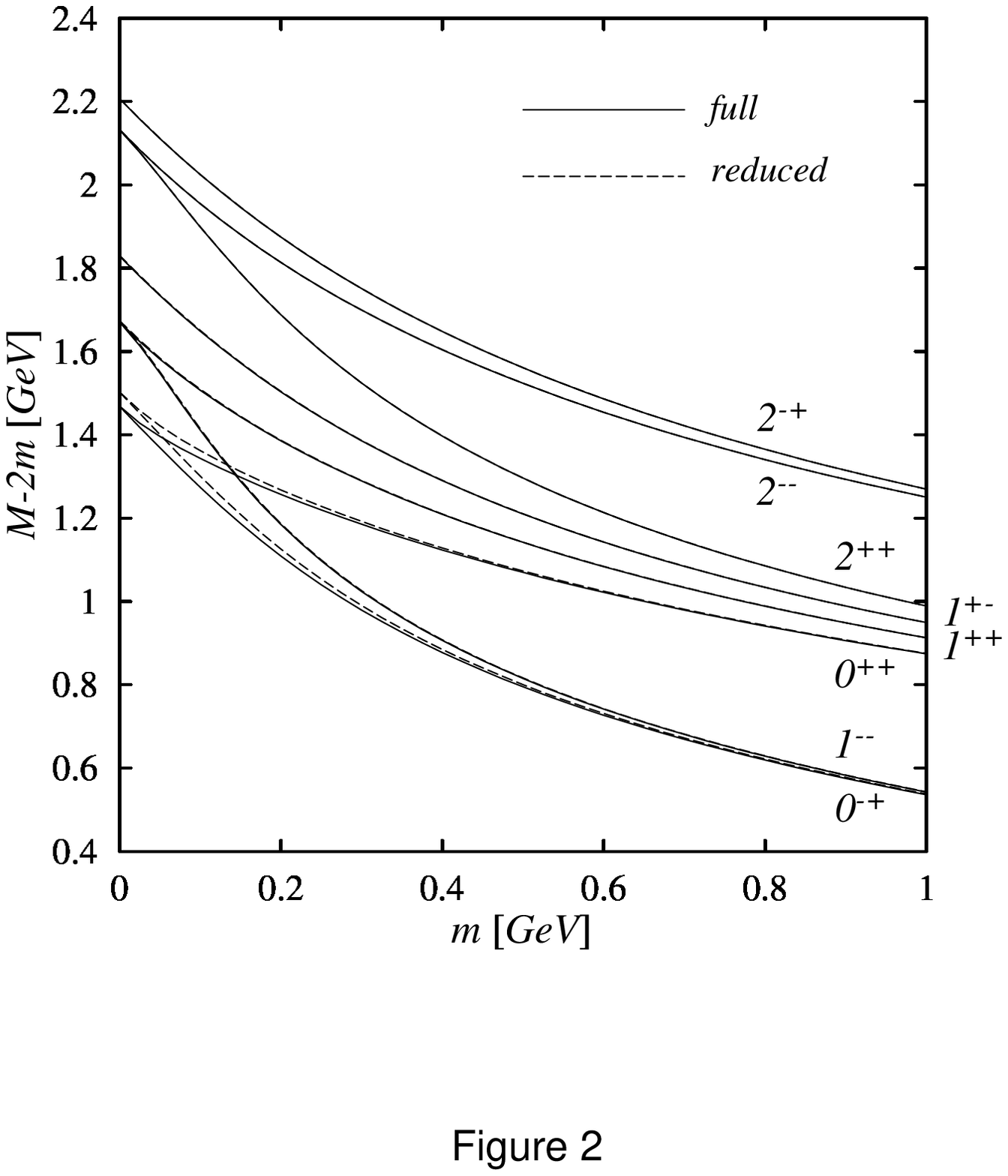}
\end{figure}

\begin{figure}[p]
\epsfxsize = 5.4in \epsfbox{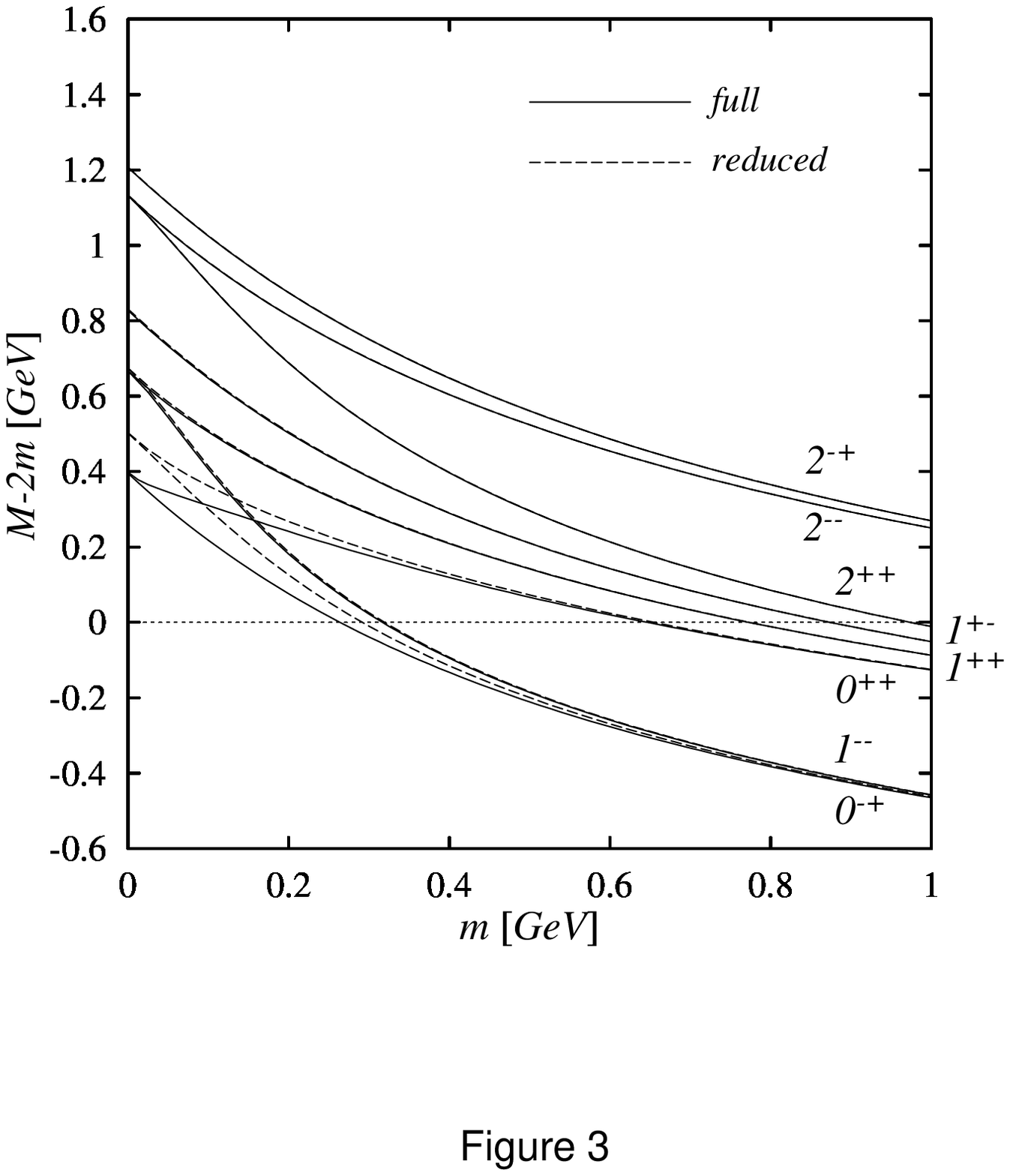}
\end{figure}

\begin{figure}[p]
\epsfxsize = 5.4in \epsfbox{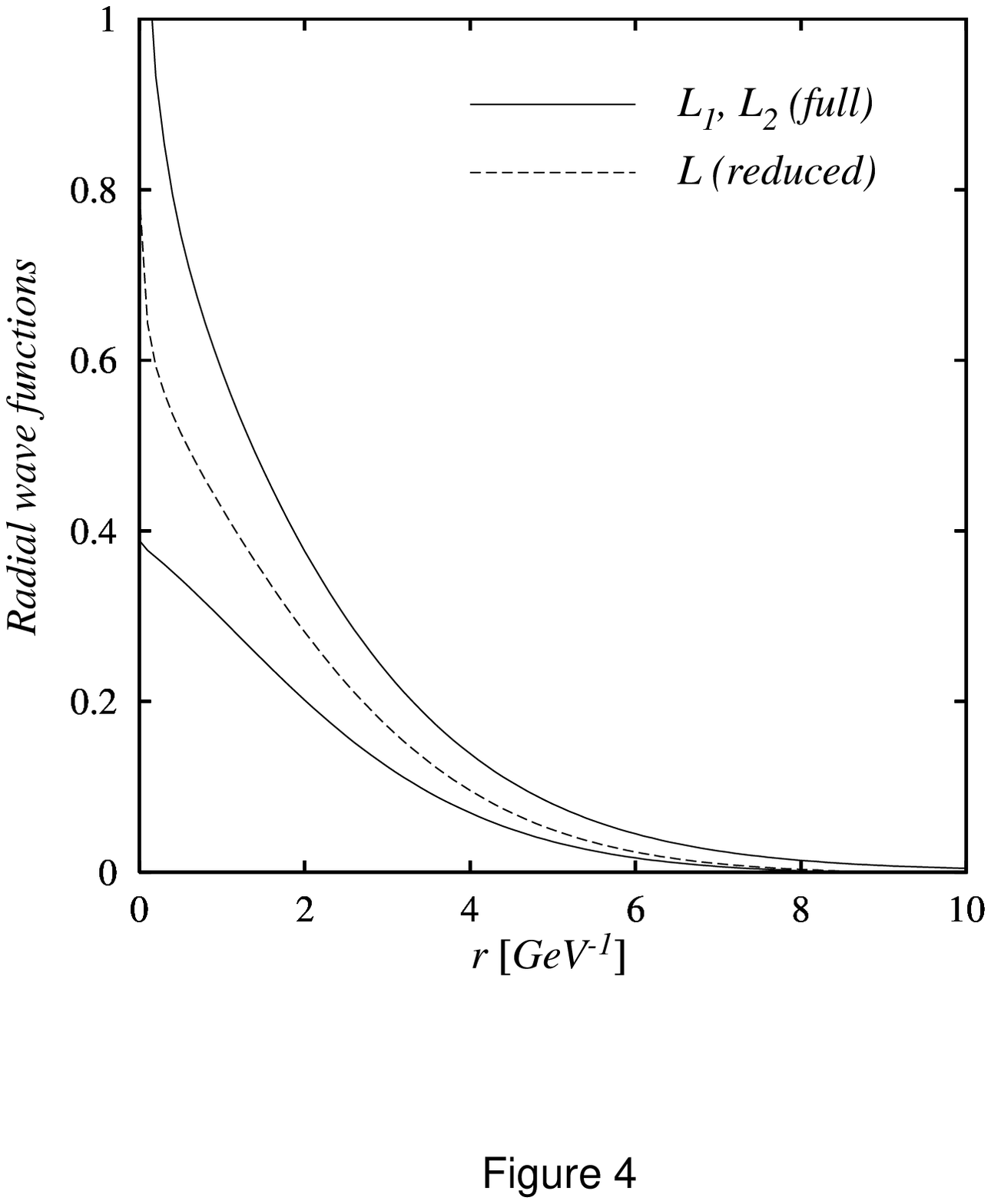}
\end{figure}

\begin{figure}[p]
\epsfxsize = 5.4in \epsfbox{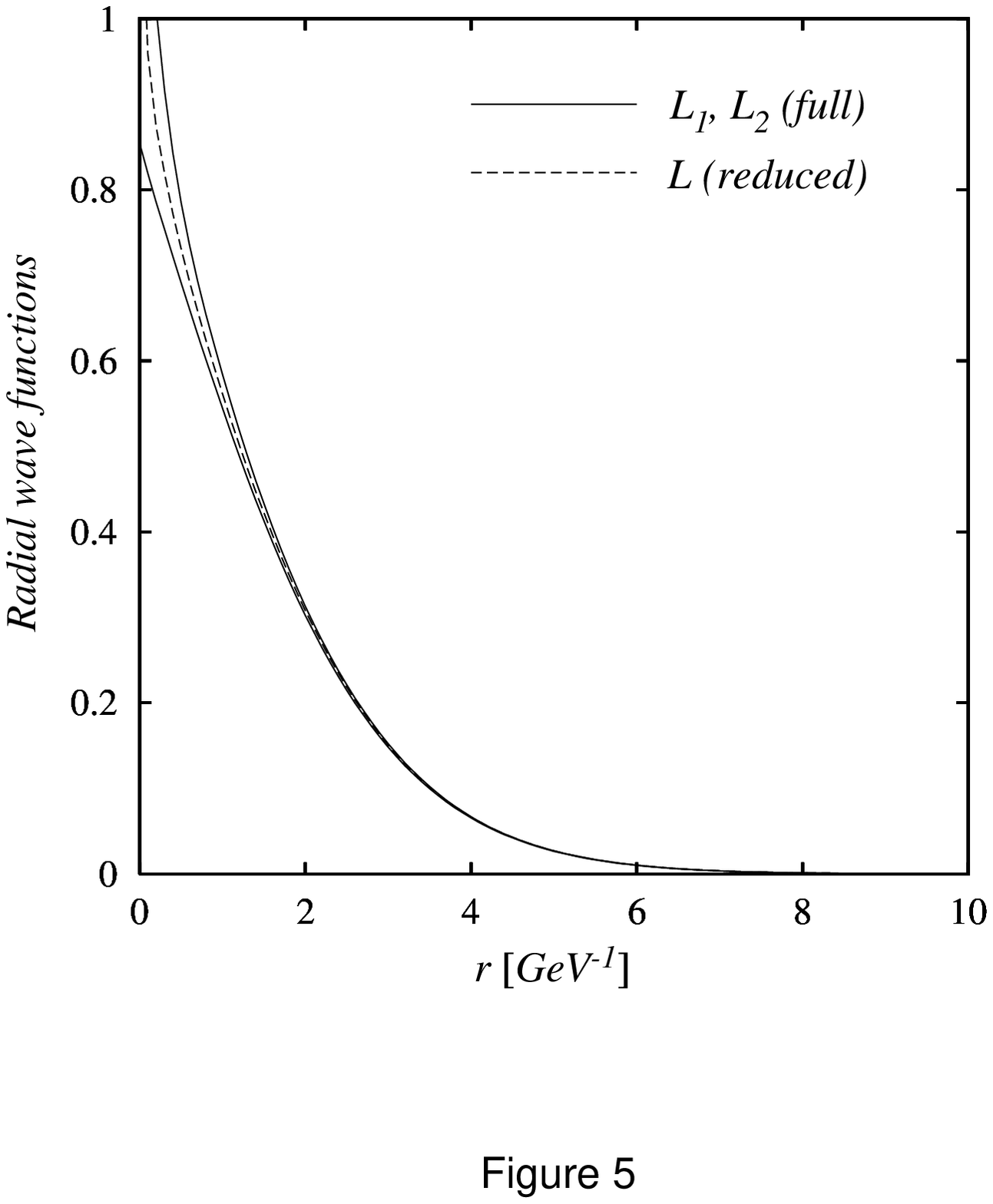}
\end{figure}

\begin{figure}[p]
\epsfxsize = 5.4in \epsfbox{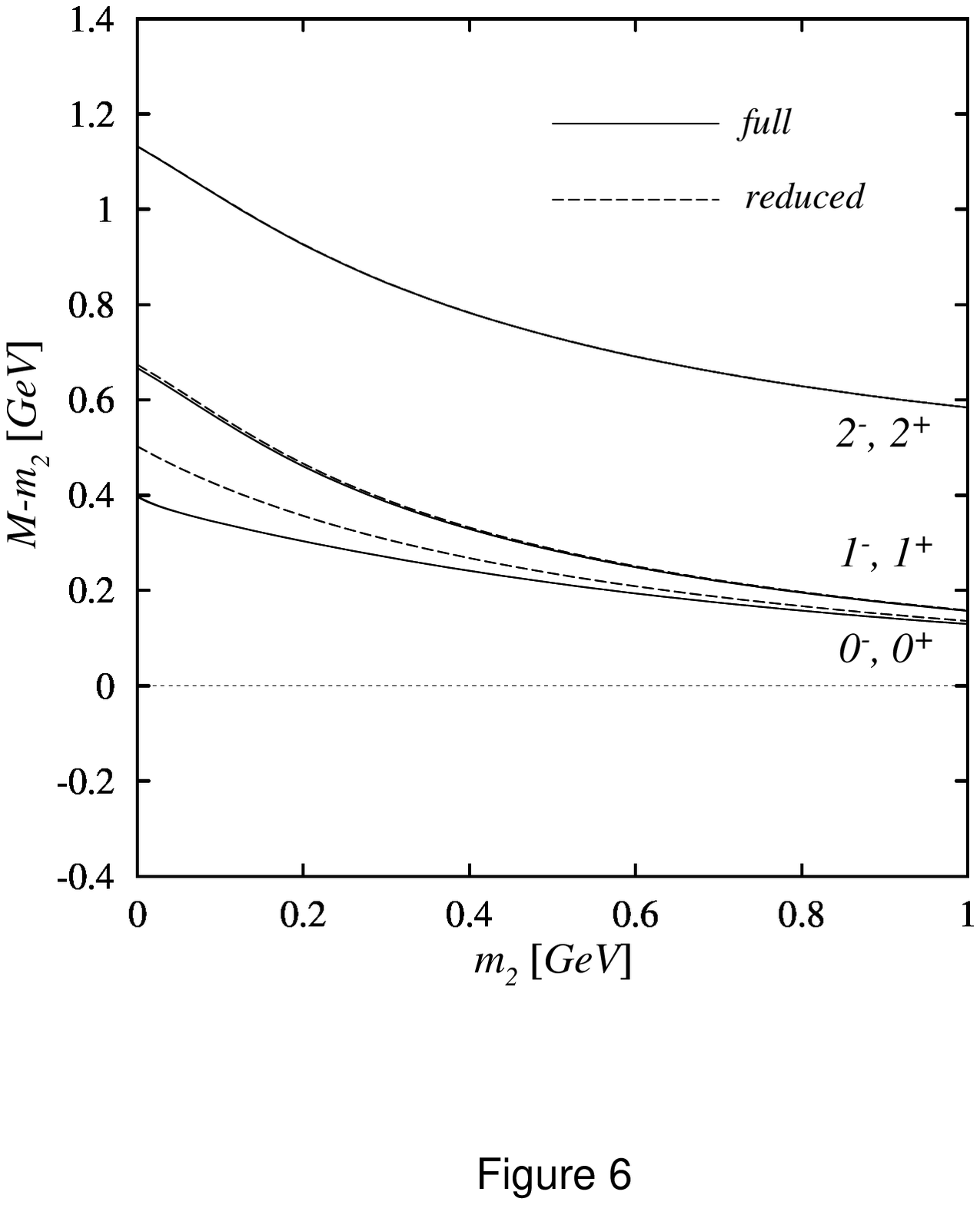}
\end{figure}

\begin{figure}[p]
\epsfxsize = 5.4in \epsfbox{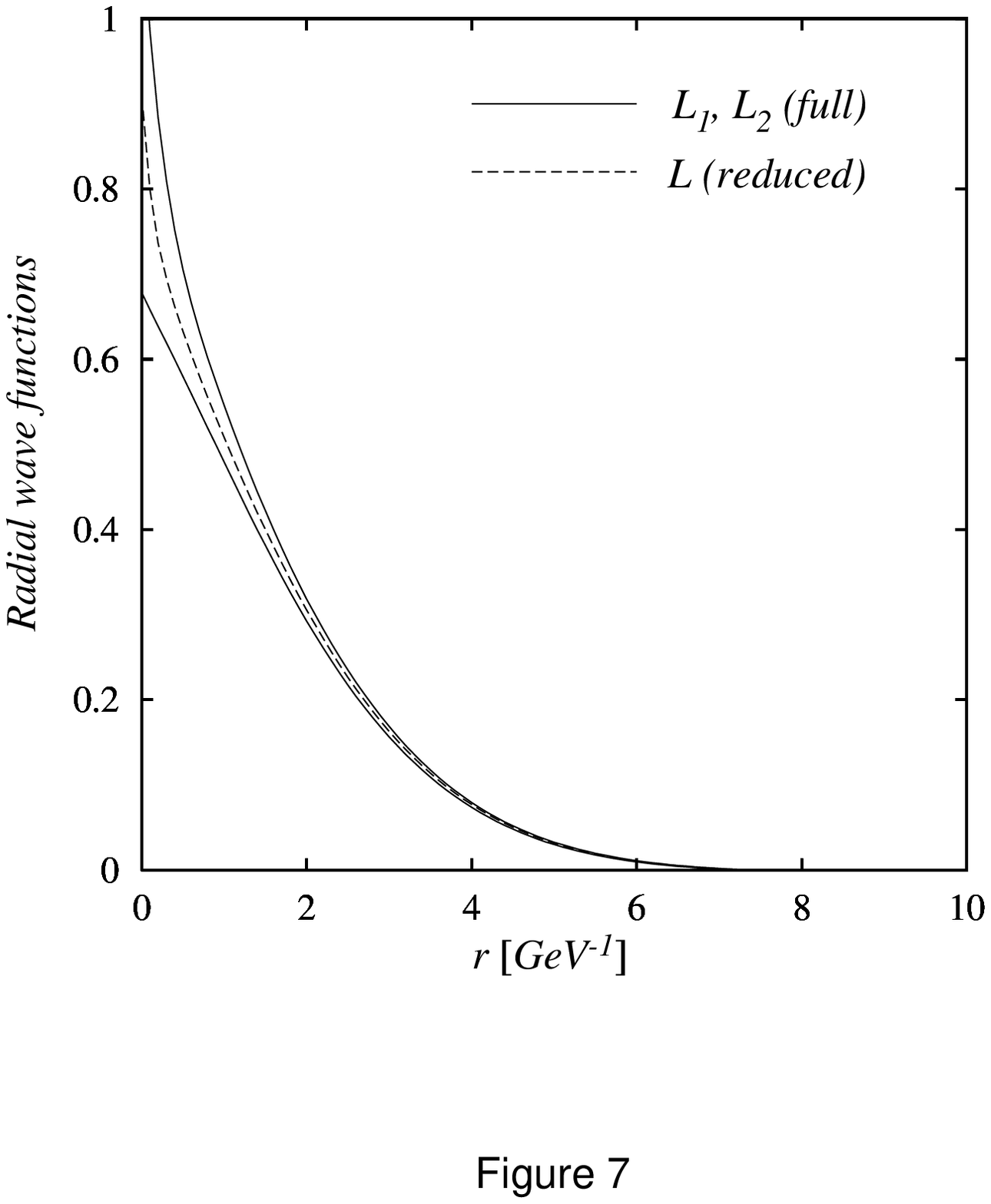}
\end{figure}

\begin{figure}[p]
\epsfxsize = 5.4in \epsfbox{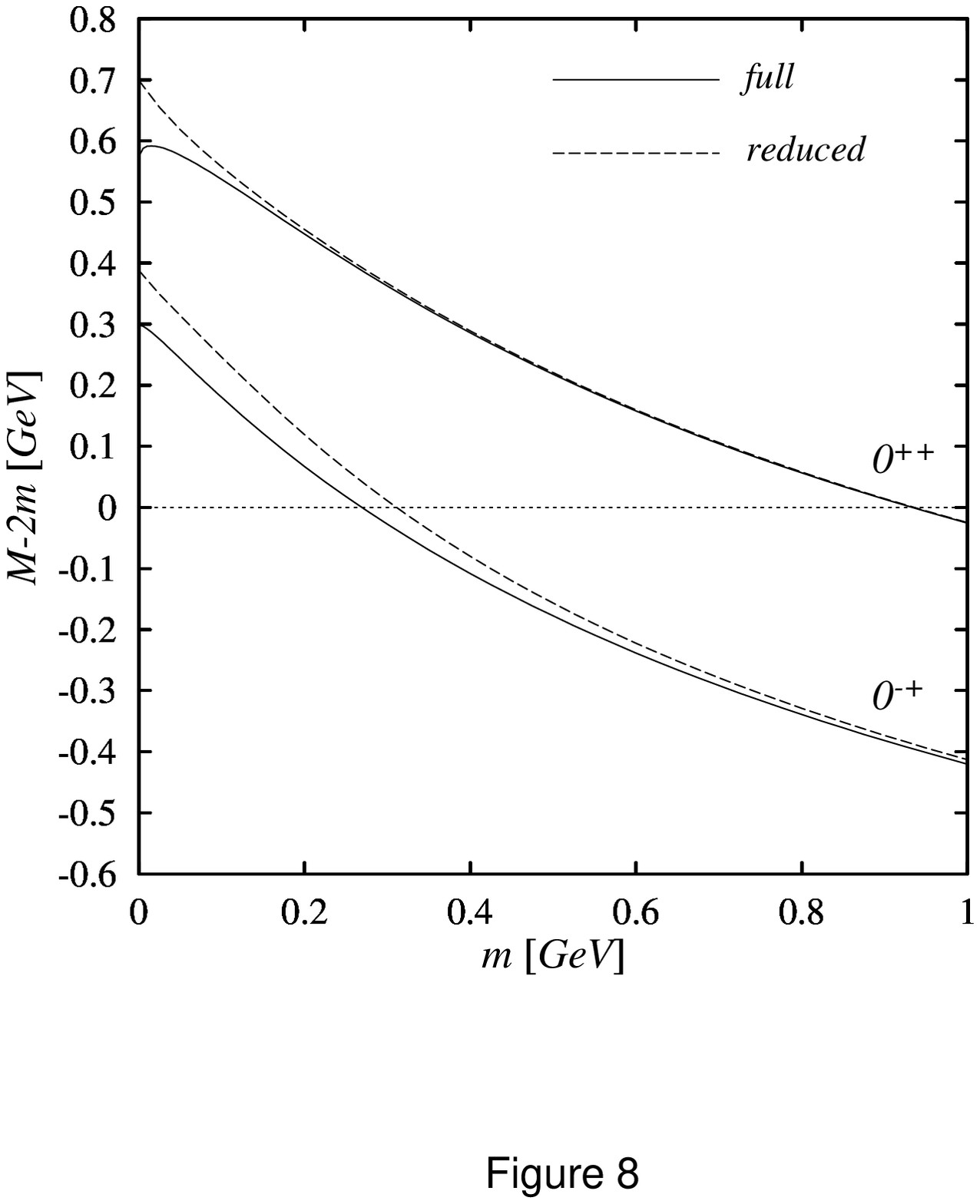}
\end{figure}

\begin{figure}[p]
\epsfxsize = 5.4in \epsfbox{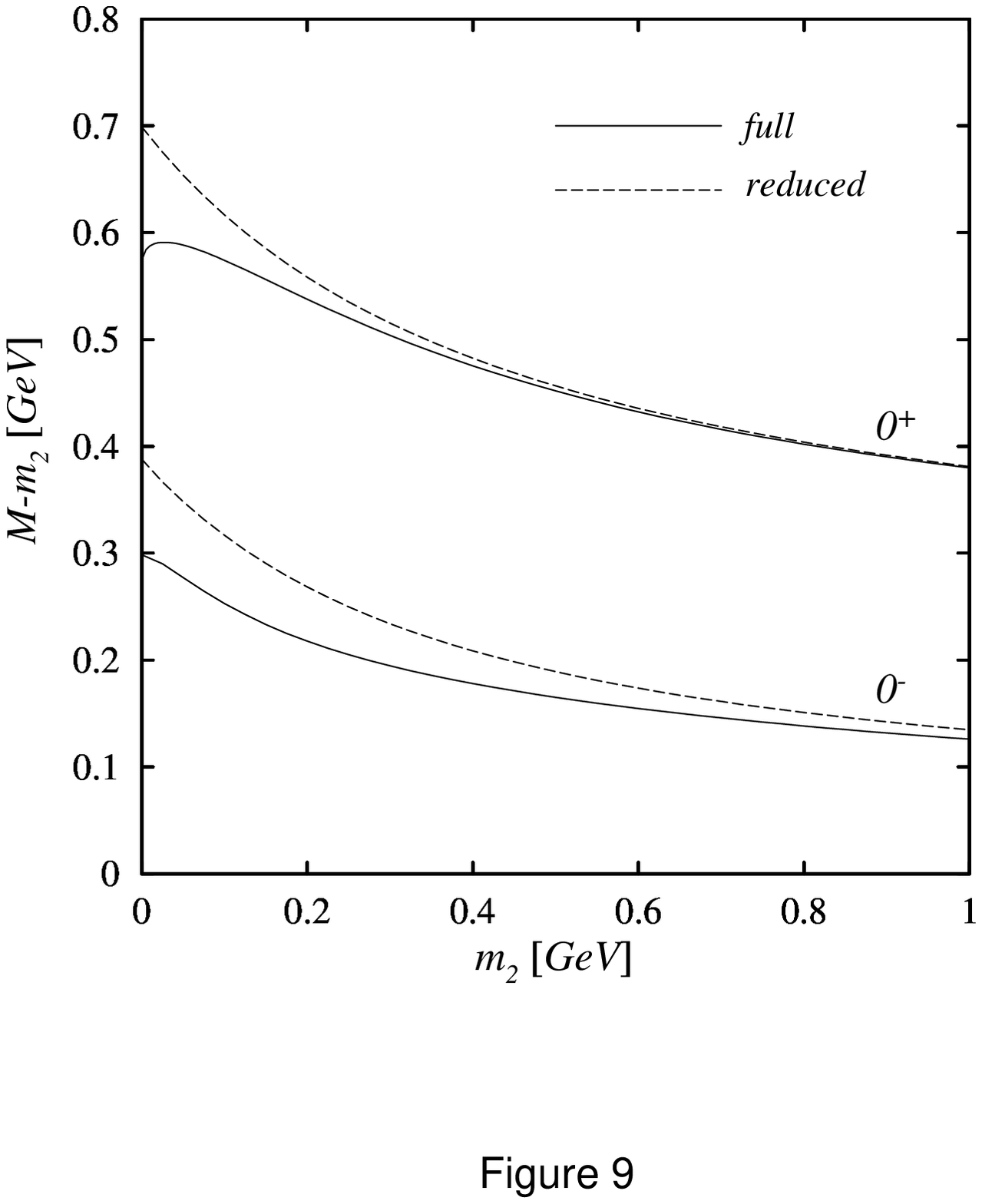}
\end{figure}

\begin{figure}[p]
\epsfxsize = 5.4in \epsfbox{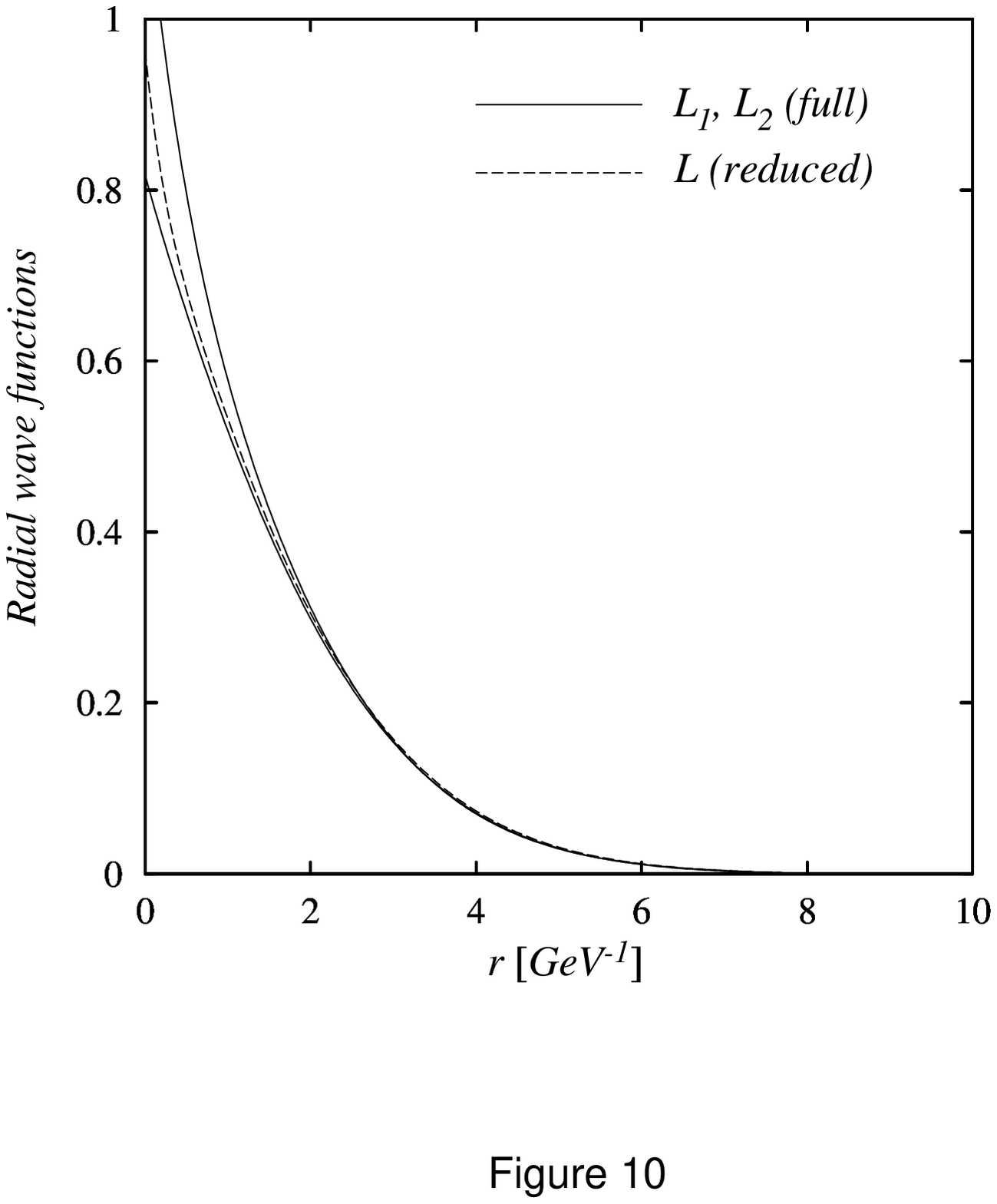}
\end{figure}

\begin{figure}[p]
\epsfxsize = 5.4in \epsfbox{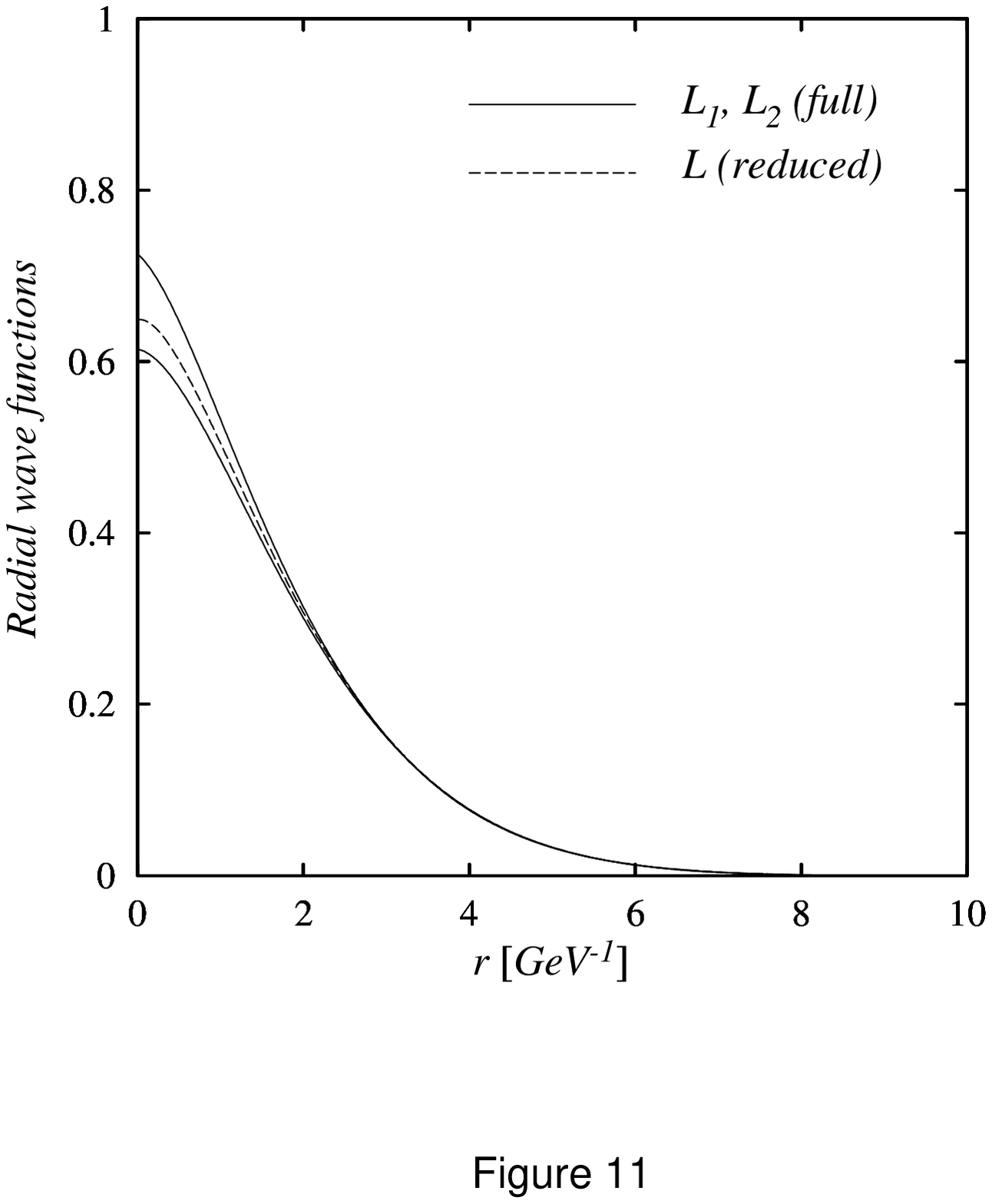}
\end{figure}

\begin{figure}[p]
\epsfxsize = 5.4in \epsfbox{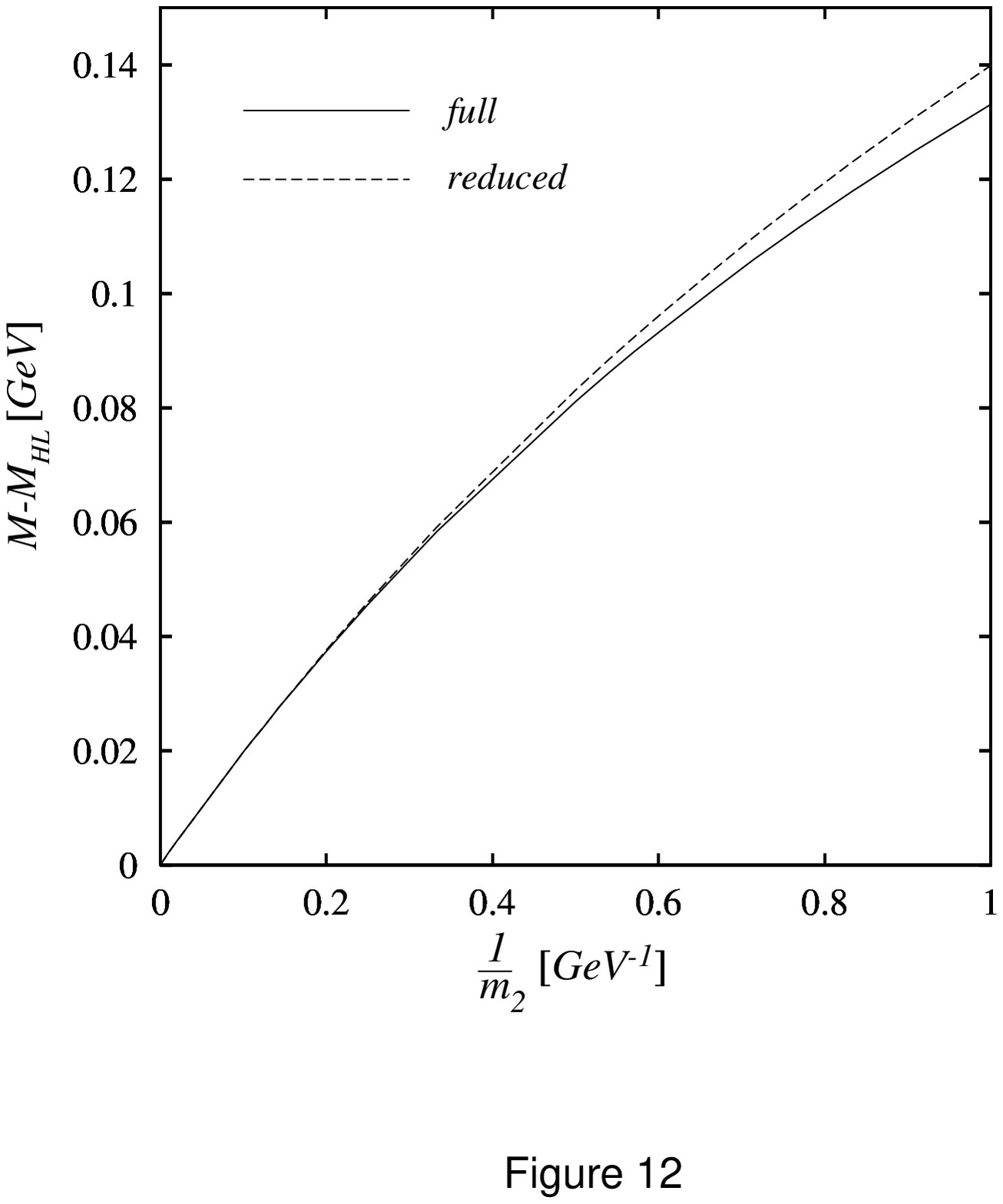}
\end{figure}


\begin{thebibliography}{99}
\bibitem{bib:lagae3} For the extensive discussion of
generic wave functions
and identification of the quantum numbers of the bound states the reader
is again referred to \cite{bib:lagae1}.
\end{thebibliography}
\end{document}